\documentclass{aa}  

\usepackage{graphicx}
\usepackage{hyperref}
\hypersetup{
    breaklinks=true,
    colorlinks=true,
    citecolor={blue},
    linkcolor={blue},
    urlcolor={blue}
}
\bibpunct{(}{)}{;}{a}{}{,}
\usepackage{txfonts}
\usepackage{natbib}  

\begin{document} 

    \title{Cosmic evolution of the Faraday rotation measure in the intracluster medium of galaxy clusters}
    
    \titlerunning{Faraday rotation measure in the intracluster medium of galaxy clusters}

    \author{Y.~Rappaz\inst{1},
          J.~Schober\inst{1,2},
          A.~B.~Bendre\inst{1,3},
          A.~Seta\inst{4},
          C.~Federrath\inst{4,5}
          }
    
    \institute{Institute of Physics, Laboratory of Astrophysics, \'Ecole Polytechnique F\'ed\'erale de Lausanne (EPFL), 1290 Sauverny, Switzerland\\
    \email{yoan.rappaz@epfl.ch}
    \and
    Argelander Institute für Astronomie, Universität Bonn, Auf dem Hügel 71, D-53121 Bonn, Germany
    \and
    Scuola Normale Superiore, Piazza dei Cavalieri, 7, 56126 Pisa, Italy
    \and 
    Research School of Astronomy and Astrophysics, Australian National University, Canberra, ACT 2611, Australia
    \and
    Australian Research Council Centre of Excellence in All Sky Astrophysics (ASTRO3D), Canberra, ACT 2611, Australia
    }

    \abstract
    {Radio observations have revealed magnetic fields in the intracluster medium (ICM) of galaxy clusters, 
    and their energy density is nearly 
    in equipartition with the turbulent kinetic energy. 
    This suggests magnetic field amplification by dynamo processes during cluster formation. However, observations are limited to redshifts $z \lesssim 0.7$, and the weakly collisional nature of the ICM complicates studying magnetic field evolution at higher redshifts through theoretical models and simulations.}
    {Using a model of the weakly collisional dynamo, we modelled 
    the evolution of the Faraday rotation measure (RM) in galaxy clusters of different masses, up to $z \simeq 1.5$, and investigated its properties such as its radial distribution up to the virial radius $r_{200}$. We compared our results with radio observations of various galaxy clusters.}
    {We used merger trees generated by the modified GALFORM algorithm to track the evolution of plasma quantities during galaxy cluster formation. Assuming the magnetic field remains in equipartition with the turbulent velocity field, we generated RM maps to study their properties.}
    {We find that both the standard deviation of RM, $\sigma_{\rm RM}$, and the absolute average $|\mu_{\rm RM}|$ increase with cluster mass. Due to redshift dilution, RM values for a fixed cluster mass remain nearly constant between $z=0$ and $z=1.5$.
    For $r/r_{200} \gtrsim 0.4$, $\sigma_{\rm RM}$ does not vary significantly with $\mathcal{L}/r_{200}$,
    with $\mathcal{L}$ being the size of the observed RM patch.
    Below this limit, $\sigma_{\rm RM}$ increases as $\mathcal{L}$ decreases.
    We find that radial
    RM profiles have a consistent shape, proportional to 
    $10^{-1.2(r/r_{200})}$, 
    and are nearly independent of redshift. Our $z \simeq 0$ profiles for $M_{\mathrm{clust}} = 10^{15}~\mathrm{M}_{\odot}$ match RM observations in the Coma cluster but show discrepancies with Perseus, possibly due to high gas mixing. Models for clusters with $M_{\rm clust}= 10^{13}$ and 
    $10^{15}~\mathrm{M}_{\odot}$ 
    at $z = 0$ and $z = 0.174$ align well with Fornax and A2345 data for $r/r_{200}\lesssim 0.4$. 
    Our model can 
    be useful for generating mock polarization observations for current and next-generation radio telescopes.}
{}


\maketitle

\section{Introduction}
In recent years, the interest in astrophysical magnetic fields 
has grown steadily, with a notable focus on the intracluster medium (ICM) of galaxy clusters.
Numerous observations have proven the existence of 
magnetic fields within the ICM. 
For example, \citet{kim_1991_excess_FRM} combined X-ray and radio 
emission data from a set of clusters to determine their radial 
Faraday rotation measure (RM) profile. 
The greatest RM variations were observed at the centre of 
some clusters, corresponding to a magnetic field of around 
$1~\mu\mathrm{G}$. 
The existence of magnetic fields of a few $\mu\mathrm{G}$ 
was revealed in both A400 and A2634 by \citet{eilek_2002_a400_a2634}. 
They also established that the fields are
ordered on a few tens of kiloparsec scales. 
Subsequently, \citet{govoni_2010_rm_hot_gc} investigated the RM 
in a large sample of hot galaxy clusters, revealing a patchy
structure on kiloparsec scales. 
All the observed magnetic field strengths reported in this sample 
are of the order of a few $\mu\mathrm{G}$s. Also, 
\citet{govoni_2010_rm_hot_gc} highlighted a lack of correspondence between the 
magnetic field and the ICM temperature. 
Many examples of studies of the dynamic properties of galaxy 
clusters and the properties of magnetic power spectra, using RM maps, 
can be found in the literature, such as
\citet{ensslin_vogt_2003_magpwspec_frm},  \citet{bionafede_2010_rot_measure_coma}, \citet{kuchar_2011_hydraA_magpwspec}, and \citet{stasyszyn_2019_galclust_dynamics}. 

Various scenarios have been proposed to explain the 
observation of $\mu\mathrm{G}$-strength magnetic 
fields in the ICM on a few tens of kiloparsec scales.
One idea is that such fields were amplified from 
primordial fields by a dynamo action caused by the 
turbulence created during cluster formation, notably by the
merging of dark matter halos \citep[e.g.][and references therein]{subramanian_2006, vazza_2014_bfields_ampl_galclust, donnert_2018_galclust_magampl, di_gennaro_2020}. 
Another possibility is that ICM magnetic fields stem from magnetized
outflows from starburst galaxies \citep{donnert_2009_magfield_clust_ouflows}. 
However, deciding in favour of one of these scenarios is a complicated task, 
not least because any trace of primordial fields may be absent in RM observations, 
and also because the fields produced by outflows are mixed by ICM turbulence, making them indistinguishable from primordial fields amplified by a dynamo \citep[see also][]{seta_federrath_2020_seedfields_ssh}. 
However, many aspects of the amplification process remain poorly understood. 
If we suppose the mechanism responsible for the amplification is the small-scale dynamo (SSD) \citep[see e.g.][]{brandenburg_2005_nonlin_dynamo, schober_2012_SSD, federrath_2016_magfield_ampl_plasmas}, this might be able to explain the presence of microgauss magnetic fields at redshifts as high as $z\simeq 0.7$, as noted by \citet{di_gennaro_2020}. 
Although the SSD can indeed be extremely fast in some systems \citep[e.g.][]{schober_2013_bfield_youngal}, it might be slower in the ICM. 
This is mainly because the magnetic field growth rate, in the case of the SSD, is 
approximately 
$1/T_{\mathrm{growth}} \sim (v_{\mathrm{turb}}/L_0)~\mathrm{Re}^{1/2}$ 
\citep{Schekochihin_2006_magfield_gclust}, 
where $L_0$ is the turbulence driving scale, $\mathrm{Re}$ is the 
hydrodynamic Reynolds number, and
$T_{\mathrm{growth}}$ is the characteristic timescale needed to amplify the magnetic 
energy to the level of equipartition with the turbulent kinetic energy, characterized by 
the turbulent velocity 
$v_{\mathrm{turb}}$.
In the ICM, these quantities are typically of the order of $v_{\mathrm{turb}}\sim 10^2~{\mathrm{km}/\mathrm{s}}$, $L_0 \sim 10-100~\mathrm{kpc}$, $\mathrm{Re} \sim 1$ \citep[e.g.][]{cho_2014_magfield_ICM_origin}, which leads to $T_{\mathrm{growth}}\sim 1~\mathrm{Gyr}$.
This characteristic time is of the same order as the dynamical time associated with galaxy cluster formation. 
However, this estimate does not take into account the 
non-linear dynamo phase, 
during which the amplification of the magnetic field is linear in time, rather than 
exponential \citep[e.g.][]{Schekochihin_2004_small_scale_dynamo, haugen_2004_hydromag_turb, bhat_2013_flucdynamo, seta_2020_saturation, kriel_2022_turbdynamo, brandenburg_2023_SSH}.
As a result, the time required for the magnetic field 
to reach equipartition with the turbulent velocity  
field
could be too long to explain certain observational 
features of the ICM magnetic fields. 
In particular, the typical lengthscale of the magnetic 
field measured by RM is of the order 
of 10-100 kpc,
which is many orders of magnitude larger than the
resistive scale $\ell_{\eta}$ in the ICM (for example, 
\citep{Schekochihin_2006_magfield_gclust} estimate $\ell_{\eta} \sim 10^4~\mathrm{km}$ for the plasma parameters of the Hydra A cluster). 
Hence, if the SSD is responsible for the 
amplification of magnetic fields in the ICM, this necessarily 
implies that the field has passed through a non-linear phase, 
causing the peak of the power spectrum of magnetic energy 
to move towards the
forcing scale of the turbulent velocity field. 
Furthermore, it is unclear whether the SSD 
can be effective in a medium where the Reynolds number is of the order of unity.
Such a dynamo
does indeed require the system to be turbulent enough so that the separation between the driving 
scale $L_0$
and the viscous scale $\ell_{\nu}$ enables a turbulent cascade to settle in. 
Since $\ell_{\nu}$ depends on $\mathrm{Re}$ through $\ell_{\nu} \sim \mathrm{Re}^{-3/4}L_0$ \citep[e.g.][]{malvadi_federrath_2023}, high enough values of $\mathrm{Re}$ are needed, although the exact limit is not well established.

The low collision rate of the ICM is due to a combination 
of high virial temperatures up to
$\sim 10^8~\mathrm{K}$ \citep[e.g.][and references therein]{moretti_2011_temp_clusters, wallbank_2022_xray_obs_temp} and a typical ion density 
approximately from $10^{-3}$ to 
$10^{-2}~\mathrm{cm}^{-3}$. This leads to 
ion-ion collision rates between $10^{-15}$ and 
$10^{-12}~\mathrm{s}^{-1}$ \citep[e.g.][]{Schekochihin_2006_magfield_gclust}. 
The corresponding  characteristic timescale of the collisions is 
approximately a hundredth of the typical dynamical time of a 
cluster that is of the order of a few gigayears
\citep[e.g.][]{ann_ewv_galaxy_clusters_formation}.
This implies that the magnetic 
field dynamics of the ICM cannot, in itself, be described by 
classical magnetohydrodynamics (MHD) equations \citep[e.g.][]{kulsrud_1983_MHD_description}. 
In a weakly collisional plasma, the isotropy of the system 
is lost as anisotropic pressures parallel and perpendicular to the 
local direction of the magnetic field occur \citep[e.g.][]{kulsrud_1983_MHD_description}. 
In the Braginskii limit \citep{braginskii_1965}, 
this pressure anisotropy is given by 
\begin{equation}\label{eq:evolution_bf_pressan}
	\Delta \equiv \frac{p_{\perp}-p_{\parallel}}{p} \simeq \frac{1}{\nu_{ii}}\frac{1}{B}\frac{dB}{dt}, 
\end{equation}
where $\nu_{ii}$ is the collision rate between ions, 
$p_{\perp}$ and $p_{\parallel}$ are the 
pressure component normal and along the local direction 
of the magnetic field, respectively, $B$ is the magnetic 
field strength, and $p$ is the total pressure.
Equation~\eqref{eq:evolution_bf_pressan} implies that the change in $\Delta$ is related to the change in magnetic field amplitude. 
However, when the value of $\Delta$ reaches $-2/\beta$ or 
$1/\beta$, where $\beta \equiv 8\pi n k_BT/B^2$ is the ratio of thermal-to-magnetic pressures (where $n$ is the particle density of the plasma, $k_B$ the Boltzmann constant, and $T$ the gas temperature),  the system enters a perturbed 
state and triggers the mirror or
firehose instabilities \citep[e.g.][]{CKW_1958_pinch, 
parker_1958_aniso_gas, vedenov_sagdeev_1958, gary_1992, 
southwood_kivelson_1993, hellinge_2007, kunz_scheko_stone_2014,melville_2016,st-onge_kunz_squire_scheko_20202, achikanath_2024_turbdyn_wcp}. 
These rapidly growing, Larmor-scale instabilities modify ion-ion 
collisionality of the plasma by producing small-scale magnetic 
field fluctuations over which the charged particles are scattered. 
This enhances the effective collisionality of the plasma, which in turn enhances the effective Reynolds number. Ultimately, this changes 
the growth rate of the magnetic field 
\citep[e.g.][]{Schekochihin_2006_magfield_gclust,
schekochihin_2010_firehose_gyro, mogavero_scheko_2014, santos-lima-2014, 
rincon_2015_mirror, st-onge_2018_hybrid}. 

Several cosmological simulations including magnetic field dynamics have been developed, such as the IllustrisTNG project \citep[e.g.][]{marinacci_2018_bfields_illustris}. 
Despite the progress that such simulations represent in the field, 
they cannot include the effects of pressure anisotropies, 
and studying such dynamics on 
magnetic field amplification in cosmological weakly collisional plasmas (WCPs) is still a
major challenge \citep[e.g.][]{wang_2021_turb_ICM}.
The main difficulty comes from the fact that the magnetic fluctuations created by kinetic instabilities 
in the ICM occur on the resistive scale that is $\ell_\eta \approx 10^{4}~\mathrm{km}$ 
\citep[e.g.][]{Schekochihin_2006_magfield_gclust}.
Including such effects in numerical simulations of a 
galaxy cluster is, therefore, inconceivable today. In contrast, the typical spatial resolution attained in the IllustrisTNG-50 simulations is of the order of $10^2~\mathrm{pc}$ \citep{nelson_2019_tng50},
which is approximately 11 orders of magnitude larger than the saturation scale of the kinetic instabilities. Increasing the resolution of cosmological simulations down to such a length scale is simply impossible today and in the near future. However, such
difficulty could be overcome if the exact expression of the relation $\mathrm{Re} = \mathrm{Re}_{\rm eff} = \mathrm{Re}_{\rm eff}(B)$ was known for 
all values of the magnetic field $B$ from the seed fields up 
until equipartition. 
However, such a relation is not well established for weak fields,
$B \lesssim 10^{-9}~\mathrm{G}$ \citep[e.g.][and references 
therein]{st-onge_kunz_squire_scheko_20202}. 

Such a magnetic field dependence of $\mathrm{Re}_{\rm eff}$ can be modelled 
in a framework that is based on merger trees (MTs) as it has been 
proposed in \citet{rappaz_schober_2024}. 
MTs are statistical tools based on the extended Press-Schechter 
theory \citep{press_schechter_model_1974}, and allow us to 
determine the merger rate of dark matter halos during galaxy cluster 
formation. 
This theory is the core of the GALFORM model 
\citep{cole2000_galform}, which uses a Monte Carlo algorithm to 
generate MTs to study galaxy formation. 
Such an approach can predict values of various galaxy properties such as the mass-to-light ratio and star formation history. 
Subsequently, the modified GALFORM algorithm was developed by 
\citet{parkinson_2008_mod_galform}, in which the GALFORM mass function 
was modified to match that of the 
Millenium simulations \citep{springel_2005_millenium}. 
In particular, \citet{jiang_2014_test_mtrees_algo} tested the 
robustness of the modified GALFORM model by comparing properties 
such as its progenitor mass function and mass assembly history
to that from cosmological $N$-body simulations. 
Using the modified GALFORM algorithm, the effect of a modified Reynolds number on the magnetic field amplification in the ICM has been studied in \citet{rappaz_schober_2024}. 
They suggest that the amplification of a primordial magnetic 
field to equipartition strengths can
occur in just a few tens of millions of years. 
It has also been suggested that the higher the redshifts at which the dynamo starts, the higher the growth rate of the magnetic field. 
Although their study only covered redshift values below $z\lesssim 2$, 
it seems plausible that magnetic field amplification could have 
started much earlier, even at $z \gtrsim 10$, when the first 
large-scale structures began to form \citep[e.g.][and references therein]{ann_ewv_galaxy_clusters_formation}.
So, if the trend of an increasing magnetic field growth rate with redshift continues beyond $z \simeq 2$, we could envisage that magnetic field amplification was almost instantaneous, compared to the characteristic dynamic time of galaxy cluster evolution.

In this paper, we study the time evolution of RM signatures during 
galaxy cluster formation and evolution, employing 
the same MT-based method as described in 
\citet{rappaz_schober_2024}. 
Drawing on the previous findings from the explosive dynamo scenario \citep[e.g.][]{Schekochihin_2006_magfield_gclust}, in this work we assume that
the magnetic field energy density is always in equipartition 
with the turbulent kinetic energy density.
We compared the model results with a sample of 
radio observations from different cluster surveys, 
such as \citet{1989_abell_catalog} and \citet{govoni_2010_rm_hot_gc}.

This paper is organized as follows.
Our MT model is presented in 
Sect.~\ref{sec:model}. 
Section~\ref{sec:results} 
presents our results, which are discussed in 
Sect.~\ref{sec:discussions}. 
Finally, our conclusions are presented in Sect.~\ref{sec:conclusions}.

\section{Model and methods}\label{sec:model}
\subsection{Merger trees}\label{subsec:model_merger_trees}
In this work, we compute MTs with the Modified GALFORM algorithm 
\citep{parkinson_2008_mod_galform} following a similar procedure as 
described in \citet{rappaz_schober_2024}. In GALFORM, the redshift, at
which the merger tree starts, is a free parameter and we consider three 
different values, namely $z_{\mathrm{max}} = 2,3,4$.
The number of redshift points between $z_{\mathrm{max}}$ and $z=0$ is 
set to $N_z = 300$, where $N_z \delta_z = z_{\mathrm{max}}$ 
with $\delta_z$ representing the redshift resolution. 
Finally, the masses of the clusters considered at $z=0$ are 
$M_{\mathrm{clust}} = 10^{13}, 10^{14}$, and $10^{15}~\mathrm{M}_{\odot}$. 
We set the ratio between the cluster mass and the 
resolution mass $M_{\mathrm{res}}$ at the same value for all 
three configurations, 
that is,~$M_{\mathrm{res}}/M_{\mathrm{clust}} = 10^{-3}$. 
Each halo at a given redshift is assumed to be in hydrostatic equilibrium.
In total, $N_{\mathrm{tree}} = 10^3$ merger trees are 
generated for each configuration.
In this work, a major merger is defined as the merger of two or more halos, 
where the mass $M$ of at least one of 
them is 
$M/M_{\mathrm{MMS}} \geq 0.1$, 
with $M_{\mathrm{MMS}}$ being the mass 
of the most massive subhalo (MMS). 
Unlike what was done in \citet{rappaz_schober_2024}, we focus our analysis on 
each most massive subhalo at a given redshift for a given MT.
This makes it possible to study global trends in the 
evolution of magnetic field observables, while limiting 
the number of assumptions -- and the biases that can 
arise from them -- imposed on any averaging process 
of physical quantities for all subhaloes at a given redshift.

In each MMS of every MT, we compute the radial profiles of all plasma quantities of interest as follows. Firstly, we assume an NFW distribution \citep{navarro_frenk_white_1996} 
for the dark matter density of the form 
\begin{equation}
  \rho_{\mathrm{DM}}(r) = \frac{\rho_\mathrm{s} }{\frac{r_s}{r} \left(1+\frac{r}{r_s}\right)^{-2}},
\end{equation} 
where $r_s$ is the scale radius, which is related to the viral radius
$r_{200} = cr_s$ through the concentration parameter $c$.
Here, $c$ 
is calculated by assuming energy conservation at each redshift step. This method, as well as the initial conditions of $c$ in all initial subhalos at $z_{\mathrm{max}}$ are discussed in Appendix E of \citet{rappaz_schober_2024}. 
The virial radius $r_{200}$ of a given subhalo 
is defined via its mass $M_h$ in
\begin{equation}\label{eq:r200}
  M_h = \frac{4}{3}\pi r_{200}^3 \Delta_c \rho_c,
\end{equation}
where $\Delta_c = 200$ and $\rho_c = 3H^2/8\pi G$ is the critical density of the universe, and $H = H(z)$ the Hubble constant as a function of redshift. The latter is given by 
\begin{equation}
  H(z)^2 = H_0^2\left[ \Omega_{0,r}(1+z)^4+  \Omega_{0,m}(1+z)^3 +  \Omega_{0,k}(1+z)^2+  \Omega_{0,\Lambda}\right]. 
\end{equation}
For the different energy densities ($\Omega_0$) and for the Hubble constant at redshift $z=0$, we adopt the values obtained from the Planck collaboration \citep{planck_2020_cosmology2018}, namely $\Omega_{0,m} = 0.315, \Omega_{0,\Lambda} = 0.685, \Omega_{0,r}=\Omega_{0,k}=0$ and $H_0 = 67.4~\mathrm{km}~\mathrm{s}^{-1}~\mathrm{Mpc}^{-1}$.

Subsequently, we calculate the radial profiles of the 
temperature and the gas density the same way as in \citet{rappaz_schober_2024}. 
More specifically, once the concentration parameter $c$ is known in each subhalo, we employ the method used by \citet{dvorkin_Rephaeli_2011} based on 
\citet{ostriker_2005_T_and_rho}. There, assuming an NFW profile 
for the dark matter distribution, the temperature and gas density profiles are respectively
\begin{equation}
T(r) = T_0 \left[1-\frac{\Pi}{1+\Phi}\left(1-\frac{\ln\left(1+r/r_\mathrm{s}\right)}{r/r_\mathrm{s}}\right)\right]
\end{equation}
and 
\begin{equation}
\rho_\mathrm{g}(r) =
\rho_0 \left[1-\frac{\Pi}{1+\Phi}\left(1-\frac{\ln\left(1+r/r_\mathrm{s}\right)}{r/r_\mathrm{s}}\right)\right]^{\Phi}, 
\end{equation}
where 
\begin{equation}
\Pi \equiv \frac{4\pi G \rho_\mathrm{s} r_\mathrm{s}^2 \mu m_\mathrm{p}}{k_\mathrm{B}T_0},
\end{equation}
with $\mu m_\mathrm{p}$ being the mean molecular weight, $m_\mathrm{p}$ the proton mass, $G$ the gravitational constant, 
$\Phi \equiv (\Gamma-1)^{-1}$ the polytropic index, 
and $\rho_s$ the integration constant of the NFW profile of the dark matter distribution.
We adopt $\mu = 1/2$ and $\Gamma = 1.2$, and $T_0$ and $\rho_0$ are determined by integration.

\begin{figure*}
    \centering
    \includegraphics[width=\textwidth]{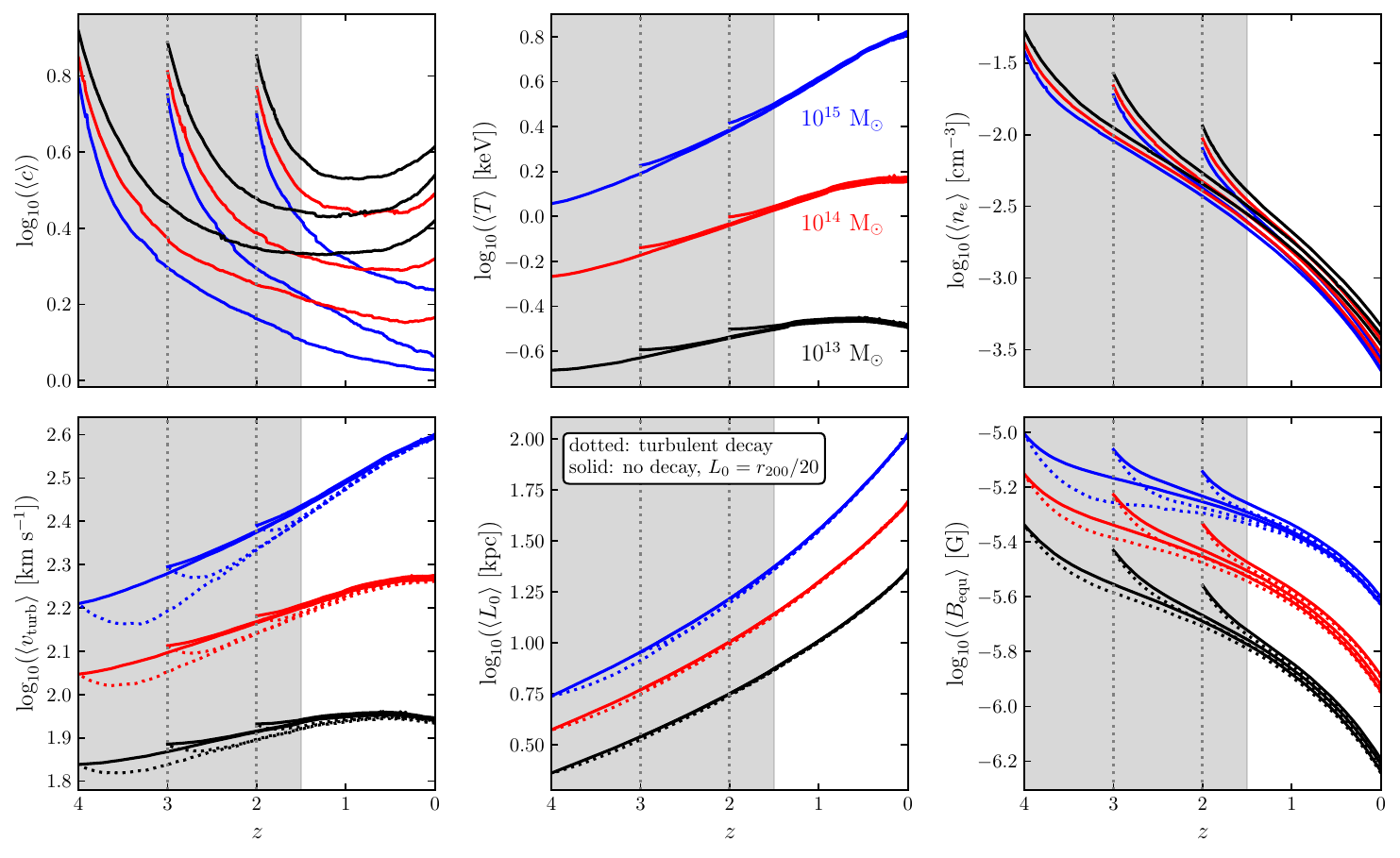}
    \caption{Evolution of the concentration parameter $c$, the temperature $T$, the electron density $n_e$, the turbulent velocity $v_{\mathrm{turb}}$, the turbulent driving scale $L_0$, 
    and the equipartition magnetic field $B_{\mathrm{equ}}$,
    as a function of the 
    redshift $z$.
    Each colour represents a different merger tree configuration with a given mass $M_{\mathrm{clust}}$ of our modelled cluster at 
    $z=0$. The mass resolution is the same in all cases, i.e. $M_{\mathrm{clust}}/M_{\mathrm{res}} = 10^{-3}$. Each merger tree mass configuration is also computed for $z_{\mathrm{max}} = 2,3,4$, where vertical grey dotted lines represent the starting point of the merging process. 
    The solid lines represent the evolution of the plasma quantities without turbulent decay.
    The coloured dotted lines correspond to the evolution of plasma quantities with turbulent decay according to power laws given by Eqs.~\eqref{eq:turb_enregy_evolution} and \eqref{eq:forcing_scale_evolution}.
    Each line represents, at a given redshift, the skew-normal median of $N_{\mathrm{tree}} = 10^3$ merger trees.
    The grey area corresponds to the part of the merger 
    trees we neglect in the subsequent analysis.}
    \label{fig:plasma_params_evolution}
\end{figure*}

\subsection{Turbulence model}
Turbulent velocity profiles have been calculated in the same way as in \citet{rappaz_schober_2024}. Specifically, we assume that velocity profiles are of the form
\begin{equation}
v_{\mathrm{turb}} = v_0\left(1+\frac{r}{r_{200}}\right)^{1/2}.
\end{equation}
We note that the exponent of $1/2$ is reminiscent of the scaling for supersonic turbulence \citep[e.g.][]{kritsuk_2007_supersonic_iso_turb, federrath_2010_inter_turb, federrath_2021_sonic_scale_turb}.
The constant $v_0$ is calculated according to the results obtained in \citet{vazza_roediger_2012} by imposing $E_{\mathrm{turb}}/E_{\mathrm{therm}} = 0.1$ in each subhalo, where $E_{\mathrm{turb}}$ and $E_{\mathrm{therm}}$ are the total turbulent and thermal energy of the halo, respectively. The turbulence driving scale is estimated based on the results in \citet{shi_2018}. 
Unlike the different scenarios studied in \citet{rappaz_schober_2024}, we assume only one fiducial value for the driving scale, namely $L_0 = r_{200}/20$. The evolution of $L_0$ for different MT configurations is shown in Fig.~\ref{fig:plasma_params_evolution} in the next section.

It was suggested in \citet{subramanian_2006} that the magnetic field 
is amplified during the major mergers epoch of a cluster formation, 
after which the turbulence starts to decay. 
They also suggest that turbulent decay can be countered by galactic wakes created by ICM galaxies.
However, in our merger tree models, major mergers will likely happen at low redshift, close to $z=0$ \citep[see Fig.~3 in][]{parkinson_2008_mod_galform}. 
To study the effect of decaying turbulence between 
the major mergers, 
we implement a decay model for 
the evolution of $E_{\mathrm{turb}}$ and $L_0$. 
Specifically, we follow the approach of \citet{subramanian_2006} and \citet{sur_2019_turb_bfields_galclust}, 
and assume that
\begin{equation}\label{eq:turb_enregy_evolution}
  E_{\mathrm{turb}}(t) \propto \left(\frac{t}{t_{i}}\right)^{-6/5},
\end{equation}
and 
\begin{equation}\label{eq:forcing_scale_evolution}
   L_0(t) \propto \left(\frac{t}{t_{i}}\right)^{-2/5},
\end{equation}
where $t_i$ is the time at which the decay starts. 
In practice, we assume that $t_i$ is the time at which a major 
merger occurs in the evolution of the cluster. 
Whenever a major merger occurs, the turbulence driving scale 
is reset to $r_{200}/20$.
At a given time $t$, we use Eq.~\eqref{eq:turb_enregy_evolution} to obtain a new value of $v_{\mathrm{turb}}$, through the relation 
\begin{equation}
E_{\rm turb} \equiv \int_{0}^{r_{200}} \rho_{\rm g}(r)v_{\rm turb}^2(r) ~2 \pi r^2 \mathrm{d}r.
\end{equation}

\subsection{Magnetic field strength and vector field}
As suggested in \citet{Schekochihin_2006_magfield_gclust},  \citet{mogavero_scheko_2014},  \citet{kunz_scheko_stone_2014}, \citet{Rincon_2016_colless_dynamo}, \citet{st-onge_kunz_squire_scheko_20202}, and \citet{rappaz_schober_2024}, the dynamo in the ICM 
during the formation of a cluster is explosive, in the sense that the characteristic time associated with the growth rate of the dynamo is extremely short compared to the dynamical time of a cluster. 
Therefore, we assume that, in the redshift range studied in this work, the magnetic field is always at equipartition with the turbulent velocity field.
The equipartition magnetic field is calculated with 
the (time-dependent) value of $v_{\mathrm{turb}}$ as 
$B_{\rm equ} = (4 \pi \rho v_{\mathrm{turb}}^2)^{1/2}$.
Assuming spherical symmetry, the radial distribution 
of the electron density and magnetic 
field is interpolated and mapped onto a three-dimensional 
matrix with the new grid cell size 
that corresponds to the turbulent forcing scale $L_0$. 
This implies that the resolution of a box with size $(2r_{200})^3$ is 
$N_{\rm cell}^3 = (2r_{200}/L_0)^3 = 40^3$, where we used the default value of the 
forcing scale $L_0=r_{200}/20$.
This choice of resolution is motivated by the SSD theory,  
where the peak of the magnetic energy spectrum moves to smaller 
wavenumbers after equipartition with the kinetic energy on the 
dissipative scale is reached. 
During this non-linear SSD phase, the minimum possible wavenumber that the
peak of the magnetic energy spectrum can reach is determined by the inverse of the turbulent forcing scale.
However, it should be noted that although it has been suggested that the dynamo of a WCP resembles that of the SSD in classical MHD with high magnetic Prandtl numbers \citep{st-onge_kunz_squire_scheko_20202}, this tendency has not yet been observed in WCP simulations \citep[see the discussion in][]{rincon_2019_dynamothies}. 

In our model, we create a 3D matrix for each component
$B_i$ with $i = x,y,z$ 
of the magnetic field vector as follows. 
We start by creating a random vector potential $\boldsymbol{A}$, 
with the components $A_i, i= x,y,z$ which are
generated randomly between $-1$ and $1$. 
Therefore, the resulting vector $\boldsymbol{F} \equiv \nabla \times \boldsymbol{A}$ yields $\nabla \cdot \boldsymbol{F}=0$. 
In a given numerical cell, if $|\boldsymbol{F}| = F$, and 
we set $|\boldsymbol{B}| = B_{\mathrm{equ}}$, 
then each component of our 3D magnetic field 
is $B_i \equiv B_{\mathrm{equ}}F_{i}/F$. 
That way, we obtain three components of a divergence-free vector 
with an amplitude of $B_{\mathrm{equ}}$.

\subsection{Rotation measure maps}
The RM 
quantifies the rotation of the polarization angle of a linearly polarized wave
as it travels through a 
magnetized 
medium \citep{burn_1966_faraday_dispersion, brentjens_2005_FRMS, ferriere_2021_frm}. 
If $\Psi_i$ is the intrinsic polarization angle of the wave, 
the final angle $\Psi_f$ after propagating through the magnetized medium is given by 
\begin{equation}\label{eq:rotated_angle}
\Psi_{\mathrm{f}} = \Psi_{\mathrm{i}} + \lambda^2\mathrm{RM},
\end{equation}
where $\lambda$ is the observing wavelength, and 
\begin{equation}\label{equ:rotation_measure}
  \begin{split}
      \frac{\mathrm{RM}}{\mathrm{rad}~\mathrm{m}^{-2}} &= \frac{e^3}{2\pi m_e^2 c^4}\int_{0}^{\mathcal{D}}\frac{n_e(s) B_{\parallel}(s)}{(1+z)^{2}} {\rm d}s\\
      &\approx 0.81 \int_{0}^{\mathcal{D}}\frac{1}{(1+z)^{2}}\left(\frac{n_e(s)}{\mathrm{cm}^{-3}}\right)\left(\frac{B_{\parallel}(s)}{\mu\mathrm{G}}\right) \left(\frac{{\rm d}s}{\mathrm{pc}}\right)
  \end{split}
\end{equation}
with $e, m_e$, and $c$, respectively, denoting the electric charge, the electron mass, and the speed of light.
In Eq.~\eqref{eq:rotated_angle}, $n_e$ is the electron density, 
$B_{\parallel}$ is the component of the magnetic field vector $\boldsymbol{B}$ 
parallel to the line of sight, and $\mathcal{D}$ is the distance from the observer to the source. The quantity $\mathrm{RM}$ is called the RM. Then, the RM is numerically computed as
\begin{equation}\label{eq:rm_numerical}
  \frac{\mathrm{RM}}{\mathrm{rad}~\mathrm{m}^{-2}} \approx \frac{0.81}{(1+z)^2}\sum_{i = 0}^{N_{\mathrm{cell}}} \frac{n_e(i)}{\mathrm{cm}^{-3}} \frac{B_{\parallel}(i)}{\mu\mathrm{G}} \frac{\Delta l}{\mathrm{pc}},
\end{equation}
where the sum is performed over all $N_{\mathrm{cell}}^2$ lines of sight of a given subhalo, and $\Delta l \equiv 2r_{200}/N_{\mathrm{cell}}$ is the size of each cell. Note that the electron density $n_e$ has also been mapped onto a 3D matrix, 
following the equipartition magnetic field. 
For the rest of this work, we set $B_{\parallel} \equiv B_{x}$. 
Let us note that every quantity computed in our merger trees is comoving. 
Therefore, adding the redshift-dependency 
to Eqs.~\eqref{equ:rotation_measure} and \eqref{eq:rm_numerical} means that we are computing the physical RM. 

\subsection{Averaging processes}
Once the radial profiles of the different plasma quantities have been calculated for each MMS in each MT, we average them in two steps to establish a redshift-dependent 1D profile.
First, we calculate the root mean square of each quantity $\Theta$ in the MMS at a given redshift, as
\begin{equation}
\langle \Theta \rangle_{\mathrm{rms}} = \sqrt{\frac{1}{N_{\mathrm{rad}}} \sum_{i=1}^{N_{\mathrm{rad}}}\Theta_i^2},
\end{equation}
where $N_{\mathrm{rad}}=10^3$ is the number of grid cells used to create every radial profile. 
A unique profile is thus obtained for each merger tree.
Then, to provide an average estimate of a physical quantity for all 
merger trees at a given redshift, we follow the same 
procedure using a skew-normal distribution 
as described in 
Appendix~C of \citet{rappaz_schober_2024}. 
In particular, the median value of such a distribution is used as the average estimator. For the rest of this work, such an average is denoted by $\langle \rangle$.
In Appendix~\ref{appendix:different_avgs} 
we discuss the effects of considering other types of averages, like a Gaussian fit or the root mean square. We have established that the different averages considered produce minor variations, and the choice of the averaging process does not impact our conclusions.

Furthermore, Faraday rotation maps are fit with a Gaussian distribution to determine their 
standard deviation $\sigma_{\mathrm{RM}}$. 
Finally, we compute radial profiles for the RM (that are compared with radio observations) as follows. 
We first take the absolute value of a given RM map. 
Then, we calculate the arithmetic average of every cell located at a given 
distance $r$ from the cluster centre.
The corresponding profile is denoted $|\mu_{\rm RM}(r)|$.

\section{Results}\label{sec:results}

\subsection{Plasma parameters}
In this section we present our results on the evolution of different plasma parameters as a function of redshift. This allows us to determine whether the values obtained for these parameters are similar to those expected from observations, numerical simulations, and other theoretical models. 
More specifically, the results in this subsection aim to test and ascertain the robustness of our approach.

Figure~\ref{fig:plasma_params_evolution} shows the evolution of 
different plasma quantities and for different merger tree configurations. 
The results of the model that assumes turbulent decay are shown as dotted lines.
The solid lines correspond to our fiducial turbulent velocity model with $L_0 = r_{200}/20$ and no decay.
First, it appears that the turbulent velocity $v_{\mathrm{turb}}$, the
driving scale $L_0$, and the equipartition field $B_{\mathrm{equ}}$ vary very little when turbulent decay is considered.
Indeed, the most significant disparity 
between the evolution of $B_{\rm equ}$ and its decaying counterpart occurs at $z \simeq 3.1$ for the model with $M_{\rm clust} = 10^{15}~\mathrm{M}_{\odot}$, where the difference between the two curves is approximately 
a factor $0.8$. This difference decreases with the 
cluster mass. 
We also note that the decay is more important at the start of the merging process.
This can be explained by the increase in major mergers over time \citep{parkinson_2008_mod_galform}. 
Since the quantities 
that are affected by turbulent decay 
only start to decrease from a major merger, the decay becomes less significant when $z\to 0.$
Furthermore, the evolution of the various curves at redshifts close to $z_{\mathrm{max}}$
is due to our initial conditions and the evolution of the concentration parameter\footnote{More precisely, we mentioned in 
Sect.~\ref{subsec:model_merger_trees} that $c$ is calculated 
by energy conservation before and after a merger. When a halo does not participate in any merger, as is often the case for redshifts close to $z_{\rm max}$, only the energy component of the accreted matter is added to the halo's energy, decreasing the value of $c$. On the other hand, as $z$ approaches zero, more mergers take place, and the value of $c$ increases. The evolution of gas density $\rho_g$ (and therefore of $n_e$) then follows a similar trend. Subsequently, as all other plasma quantities are calculated as a function of $n_e, T,$ and $v_{\rm turb}$, they also 
follow a similar trend.}, which encourages
us to focus particularly on the effects occurring at low redshift. 
Considering this, we can establish that 
the modelling of turbulent decay laws
will not have a 
significant influence on our conclusions, and are not considered in the rest of this work.
Moreover, the parameter $z_{\mathrm{max}}$ has 
minimal influence on the curves as they approach $z = 0$.
Conversely, the trajectories of the various curves in proximity to $z_{\mathrm{max}}$ exhibit a similar overall trend. This comes mainly from the evolution of the concentration parameter $c$, which is similar in 
models with different $z_{\rm max}$, as it is displayed in 
Fig.~\ref{fig:plasma_params_evolution}. Indeed, the
concentration parameter is calculated by energy conservation between two redshift steps, whether or not a merger occurs. Furthermore, the accretion component is also taken into account in the energy \citep[see Appendix E of][]{rappaz_schober_2024}. Generally, mass accretion is more important at high redshift, which would explain such a trend.

To ensure methodological rigor, our investigation will focus exclusively on redshift values ranging 
from 0 to 1.5, mitigating the potential impact of numerical or model artefacts on the results. Additionally, we will exclusively focus on the configuration with $z_{\mathrm{max}}=4$ to prevent any redundancy in the outcomes.

From Fig.~\ref{fig:plasma_params_evolution}, 
we see that 
the temperatures obtained at $z=0$ for $M_{\mathrm{clust}} = 10^{14}$ and $10^{15}~\mathrm{M}_{\odot}$ are approximately between 1.5 and 7~keV. Such values are well in line with the results reported from various X-ray surveys \citep[see e.g.][and references therein]{moretti_2011_temp_clusters, baldi_2012, wallbank_2022_xray_obs_temp}. 
Although our model does not replicate the characteristics of cool-core clusters, the temperatures mentioned above (from our model and observations) are averaged quantities, allowing us to compare them regardless of the
dynamic state of the sample of observed clusters.

Also, the typical turbulent velocities from our model are of the order of $\sim 150~\mathrm{km}~\mathrm{s}^{-1}$ for $M_ {\rm clust} = 10^{14}~\mathrm{M}_{\odot}$ 
and
$\sim 300~\mathrm{km}~\mathrm{s}^{-1}$ for $M_ {\rm clust} = 10^{15}~\mathrm{M}_{\odot}$. Such values are of the same order of magnitude as the ones derived from X-ray observations of a sample of cool-core clusters
obtained by Chandra,
as presented in \citet{zhuravleva_2018_turbulence_cool_cores}, 
which vary between approximately 100 and 150
$\mathrm{km}~\mathrm{s}^{-1}$.
Generally, cool-core clusters are dynamically relaxed. Moreover, sufficient time must elapse after a major merger for the system to reach such a state. 
It is therefore natural to wonder if our comparison with observations of cool-core clusters is 
consistent.
In fact, in the modified GALFORM model, the major mergers of lower-mass clusters tend to occur at higher redshifts. For 
$M_{\rm clust} = 10^{15}~{\rm M}_{\odot}$ 
for instance, the major mergers most often occur at a redshift of $z \simeq 0.25$. By adopting the same cosmological parameters as those described in Sect.~\ref{subsec:model_merger_trees}, this corresponds to approximately 3 Gyr elapsed since the last major merger. This is comparable to the prediction made by \citet{richardson_2022_last_majormerg} for the A2345 cluster, which estimates its last major merger to have occurred approximately 2.2 Gyr ago.
Overall, our MT-based models produce results that match low-redshift observations well.

\begin{figure*}[h!]
    \centering
  \includegraphics[height=0.9\textheight]{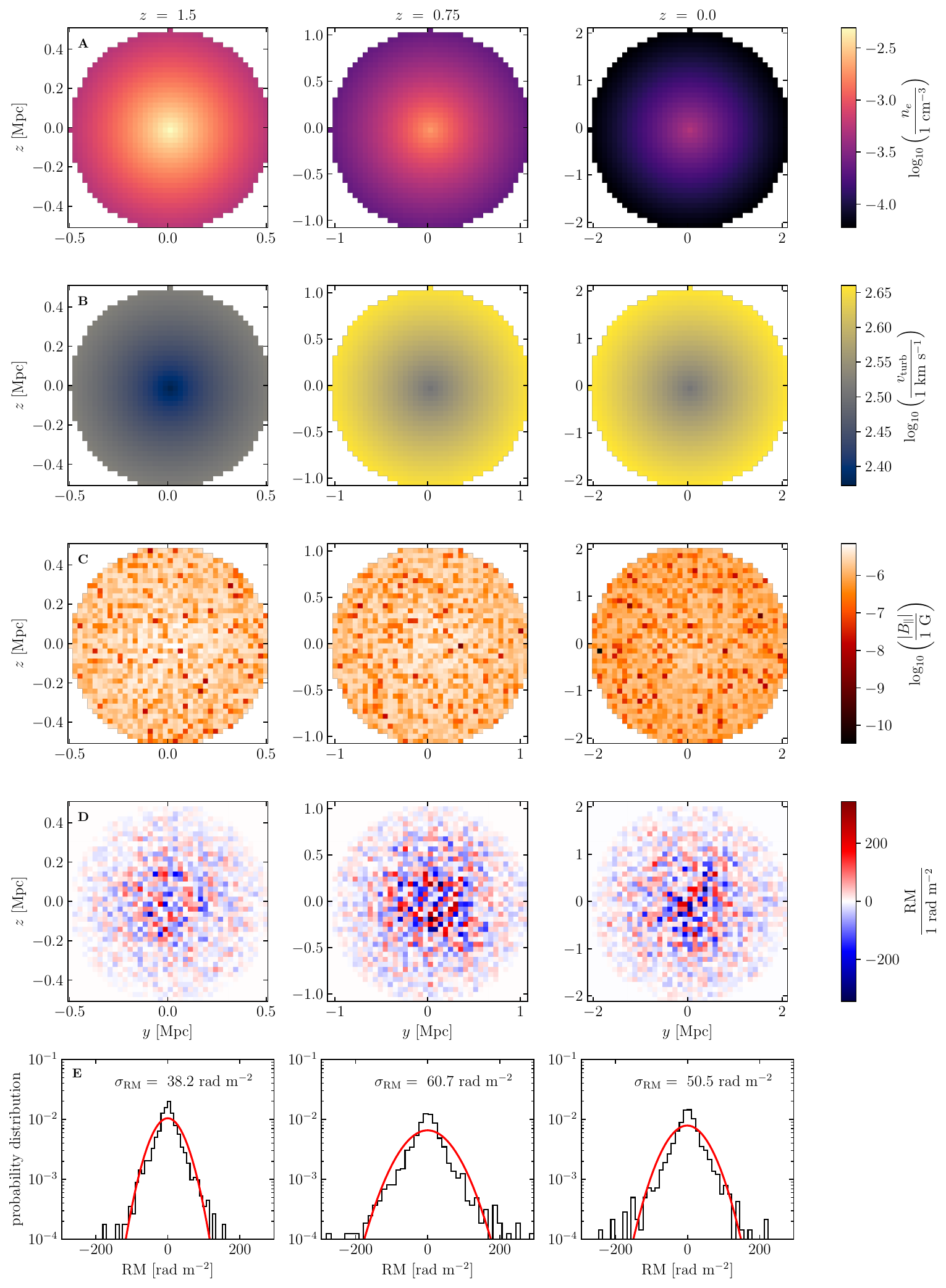}
    \caption{ 
    Spatial distribution of characteristic quantities for one realization of a merger tree. The example shows a cluster with $M_{\mathrm{clust}} = 10^{15}~\mathrm{M}_{\odot}$ at $z=0$, which has been modelled starting at $z_{\mathrm{max}} = 4$. 
    Each column corresponds to a different value of redshift, as indicated on top.
    (A) Central 2D slice of the 3D distribution of the electron density. 
    (B) Central 2D slice of the 3D distribution of the turbulent velocity. 
    (C) Central 2D slice of the 3D distribution of the line-of-sight component of the magnetic field. 
    (D) Rotation measure map. 
    (E) Probability distribution function of the RM maps. 
    The red lines correspond to a fit of the histogram data with a Gaussian distribution with the corresponding standard deviation of the Gaussian distribution, $\sigma_{\rm RM}$, shown in the legend. 
    The mean of the RM distribution is $\approx 0~{\rm rad~m}^{-2}$, as expected for the random magnetic fields.}
    \label{fig:example_rm_map_histograms}
\end{figure*}

\begin{figure}
    \centering
    \includegraphics[width=\columnwidth]{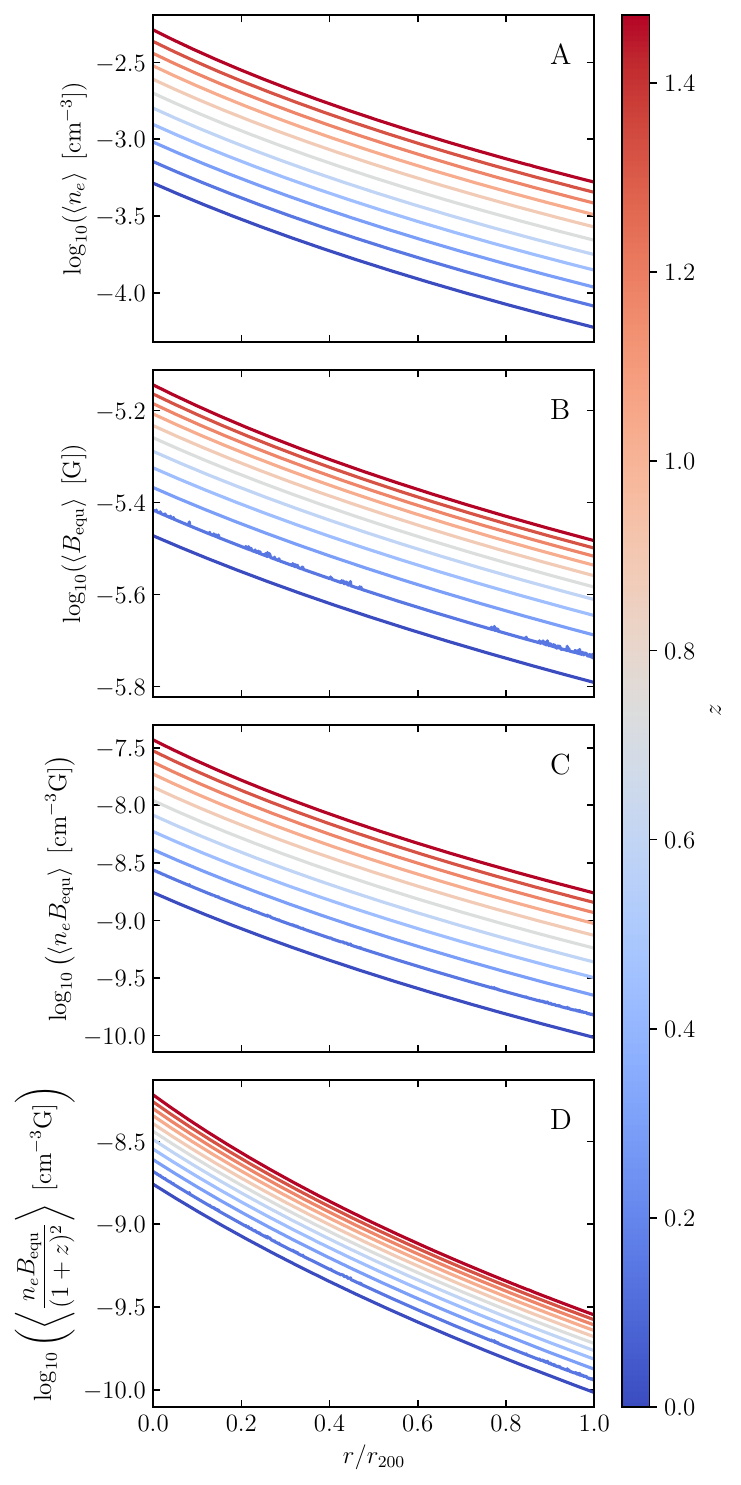}
   \caption{Radial distribution of the electron density (A), the equipartition magnetic field (B), and the product of the two with (D) and without (C) redshift dilution, for $M_{\mathrm{clust}} = 10^{15}~\mathrm{M}_{\odot}$ and $z_{\mathrm{max}} = 4$. Each value at a given radius is the skew-normal median of the values of all $N_{\mathrm{tree}} = 10^3$ MTs. We can see that the $n_e$ profiles decrease more than that of $B_{\rm equ}$. Without redshift dilution, we would expect $n_e$ to play a major role in the RM decay. However, because of the $(1+z)^{-2}$ factor, RM curves evolve very little over time.}
    \label{fig:radial_profile_ne_bequ}
\end{figure}

\subsection{Rotation measures}
In this section we present the results of the RM maps produced in our merger trees.  
We calculate $\sigma_{\rm RM}$ for regions of varying fixed spatial sizes, as radio observations typically do not encompass the entire cluster and are limited by the coverage of the sources.
We also compute the radial profiles of the RM.

\subsubsection{RM maps for $10^{15}~\mathrm{M}_{\odot}$ clusters}
Figure~\ref{fig:example_rm_map_histograms} shows an example of the RM map calculated for $M_{\mathrm{clust}} = 10^{15}~\mathrm{M}_{\odot}$, for different redshift values. 
We also show central 2D slices of the 3D distributions of the thermal electron distribution, the turbulent velocity, and the magnetic field component parallel to the line of sight, along which the RM is calculated. In addition, the distributions of the RM are fit with a Gaussian distribution, to illustrate how the standard deviation of the RM, denoted by $\sigma_{\mathrm{RM}}$, is calculated (the mean is $\approx 0~{\rm rad~m}^{-2}$, which is expected of the random magnetic fields in the cluster). 
The RM 
distribution computed from the merger tree calculation matches well with a Gaussian distribution, highlighting that the correlation between thermal electron density and magnetic fields \citep[see Fig.~C3 in][]{seta_2023} does not play a significant role \citep{seta_federrath_2021}. 

For the redshift evolution, we note that the range of RMs and the associated $\sigma_{\rm RM}$ does not
evolve significantly 
(Fig.~\ref{fig:example_rm_map_histograms}, panel E).
This can be explained, in particular, by the evolution of the radial distributions of the thermal electron density $n_e$ and
the equipartition field $B_{\rm equ}$, 
which are shown in Fig.~\ref{fig:radial_profile_ne_bequ}. 
On one hand, these profiles decrease when $z\to 0$, which is supposed to produce a corresponding decrease in RM magnitude (pannel C in the figure). On the other hand, the factor $(1+z)^{-2}$ in \eqref{eq:rm_numerical} attenuates the resulting RM at high redshift, thus decreasing its value. The redshift-diluted RM curves are then much closer together, which explains its very low variation in magnitude over time. 

\begin{figure*}
    \centering
    \includegraphics[width=\textwidth]{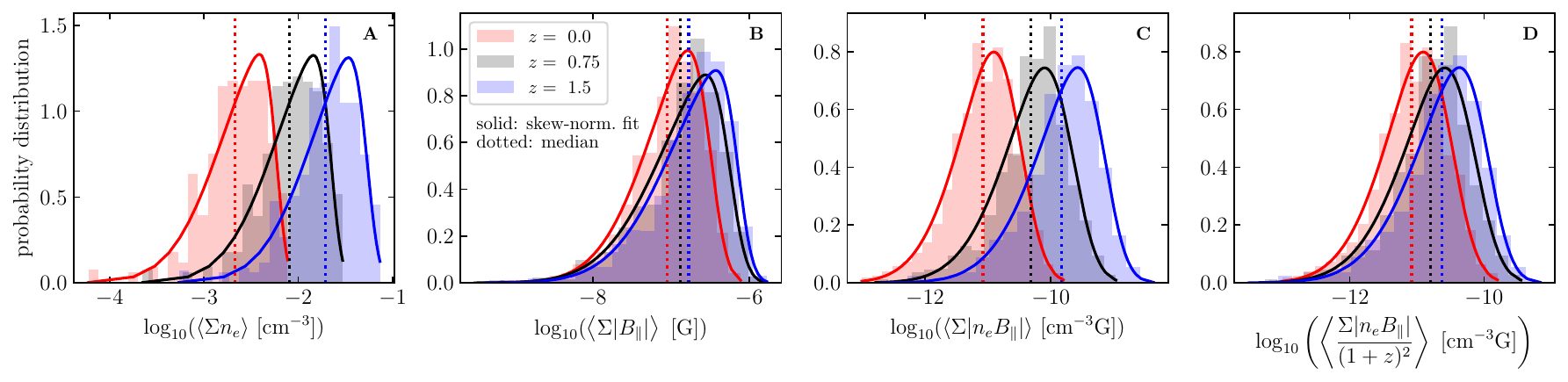}
    \caption{Average distribution of the sum along the line ($x$-axis) of sight of the electron density (A), the parallel component of the magnetic field (B), the product of those two quantities (C), and the product divided by the $(1+z)^{-2}$ contribution, for $N_{\mathrm{tree}} = 10^3$ MTs of our model with $M_{\mathrm{clust}} = 10^{15}~\mathrm{M}_{\odot}$. In the same panel, each distribution corresponds to a given redshift value. Each histogram is fitted with a skew-normal 
    distribution (see Appendix C of \citet{rappaz_schober_2024}) which is shown in solid lines and the median value of the latter is displayed in dotted lines, to highlight which parameter amongst the electron density and the line-of-sight component of the magnetic field has the most influence on the rotation measure. It appears that if the contribution of the expansion of the universe is not taken into account, the distribution of thermal electron density is dominant. With the expansion, the RM shows minor variations.}
    \label{fig:main_com_rm}
\end{figure*}

Figure~\ref{fig:main_com_rm} shows the sum along the line of sight of the maps of the electron density, $\left\langle\Sigma n_e \right \rangle$ (panel A), the parallel magnetic field component $\left\langle\Sigma B_{\parallel}\right\rangle$ (panel B), their product $\left\langle\Sigma n_eB_{\parallel}\right\rangle$ (panel C), and their product divided by $(1+z)^2$ (panel D), for three different redshift values
The maps of all $N_{\rm tree}=10^3$ MTs are fitted with a skew-normal distribution\footnote{Taking the sum of the different quantities along the $x$-axis leads to a distribution closer to a skew-normal distribution. This is why we choose not to fit the distributions with a Gaussian law.}, and the median values of these distributions are used to approximate the trend of each curve. It appears that the mean value of the distribution of $\Sigma B_{\parallel}$ hardly changes over time, despite the overall decrease of $B_{\mathrm{equ}}$ when $z\to 0$. This is mainly because the magnetic field components are generated randomly, and even re-scaling with the amplitude of $B_{\mathrm{equ}}$ does not generate a significant difference between the distributions. On the other hand, we can see that $\left\langle\Sigma n_e\right\rangle$ and $\left\langle \Sigma n_eB_{\parallel} \right\rangle$ follow a similar pattern.
However, the sum $\left\langle\Sigma n_eB_{\parallel}/(1+z)^2\right\rangle$ seems to have the same tendency as $\left\langle\Sigma B_{\parallel}\right\rangle$. Therefore, the evolution of the thermal electron density $n_e$ coupled with the $(1+z)^{-2}$ factor tends to maintain
the observed RM at an approximately constant average value.

\begin{figure*}
    \centering
    \includegraphics[width=\textwidth]{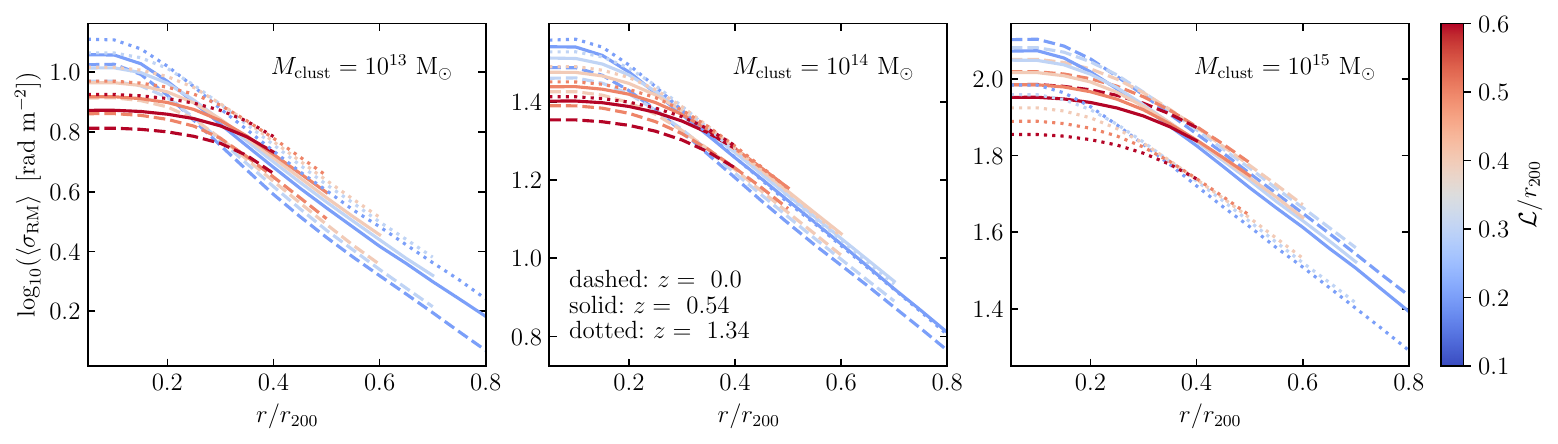}
    \caption{Evolution of the standard deviation $\sigma_{\mathrm{RM}}$ of the RM computed for the equipartition magnetic field $B_{\mathrm{equ}}$ as a function of the redshift. 
    Each column corresponds to a specific MT configuration. 
    The different colours correspond to the ratio of the typical 
    size $\mathcal{L}$ of the observed RM patch to the virial 
    radius of the halo.
    Note that the profiles are calculated up to the value of $r$ where the edge of the extracted
    sub-matrix with a size corresponding to a given value of 
    $\mathcal{L}/r_{200}$ reaches $r_{200}$. 
    Therefore, for higher values of $\mathcal{L}/r_{200}$ the profiles extend up to lower values of $r/r_{200}$.
    }
    \label{fig:rm_std}
\end{figure*}

\subsubsection{Evolution of the RM standard deviation}\label{ssubsec:rm_stddev_evol}
Figure~\ref{fig:rm_std} shows the value of the standard deviation of the RM, denoted $\sigma_{\mathrm{RM}}$, for different
MT configurations and redshift values. 
We consider different characteristic sizes $\mathcal{L}$ of observed RM patches, which are normalized by the virial radius $r_{200}$ of the dark matter halos. 
The reason for conducting such an analysis is that the RM is never observed for the entire structure of a cluster. Radio sources (in the background or inside the cluster) are limited, and the resulting RM patches often have a characteristic size much smaller than that of the observed cluster. It is therefore legitimate to study the trend of $\sigma_{\rm RM}$ as a function of the size of the observed RM maps.

For each value of $\mathcal{L}$, we extract the submatrix of the corresponding size, centred on each grid cell along the line passing through the centre of the halo. Then, $\sigma_{\mathrm{RM}}$ is calculated for all submatrices, which leads to a radial profile $\sigma_{\mathrm{RM}} = \sigma_{\mathrm{RM}}(r, \mathcal{L})$. The operation is repeated 
for each of the $N_{\mathrm{tree}} = 10^3$ MTs, and the curves are then averaged by a skew-normal distribution, 
which tends to smooth out the effect of individual variations for each halo. This can be seen, for example, by comparing Figs.~\ref{fig:rm_std} and \ref{fig:effect_incr_factors}.
Given the spherical symmetry assumed for each halo, performing the same calculations along a different axis is not expected to produce different results.

In our model, $\langle\sigma_{\mathrm{RM}}\rangle$ 
is smaller for higher values of $\mathcal{L}$, at least up to $\mathcal{L}/r_{200} \simeq 0.4$, where all curves seem to match. For example, $\mathcal{L}/r_{200}=0.1$ corresponds to a $3\times 3$ submatrix, containing only $9$ data points. The corresponding distribution is therefore not continuous, and the fitted curve by a Gaussian distribution could be wider, producing a higher standard deviation. We also see that, for $M_{\rm clust} = 10^{13}$ and $10^{14}~\mathrm{M}_{\odot}$, the distributions at different redshifts
almost coincide; however, this trend 
is not the same for $M_{\rm clust} = 10^{15}~\mathrm{M}_{\odot}$. 
This can be understood as the effect of the term $(1+z)^{-2}$ in 
\eqref{eq:rm_numerical} on the average RM magnitude, which will be also 
discussed in Sect.~\ref{ssubsec:avg_rm_profiles}. 
Overall, our model predicts that the distribution of
$\langle\sigma_{\mathrm{RM}}\rangle$ is not expected to vary significantly 
at different redshifts for $M_{\rm clust} \lesssim 10^{14}~\mathrm{M}_{\odot}$. 
However, we expect $\langle\sigma_{\mathrm{RM}}\rangle$ to decrease in magnitude 
as $z$ increases for a characteristic cluster mass somewhere in 
$10^{14}~\mathrm{M}_{\odot} \lesssim M_{\rm clust} \lesssim 10^{15}~\mathrm{M}_{\odot}$. 
Establishing this `turning-point'
mass would require more values of $M_{\rm clust}$ to be studied with our merger 
trees, which can be done in future work.

\subsubsection{Averaged RM radial profiles}\label{ssubsec:avg_rm_profiles}
\begin{figure*}[h!]
    \centering
    \includegraphics[width=\textwidth]{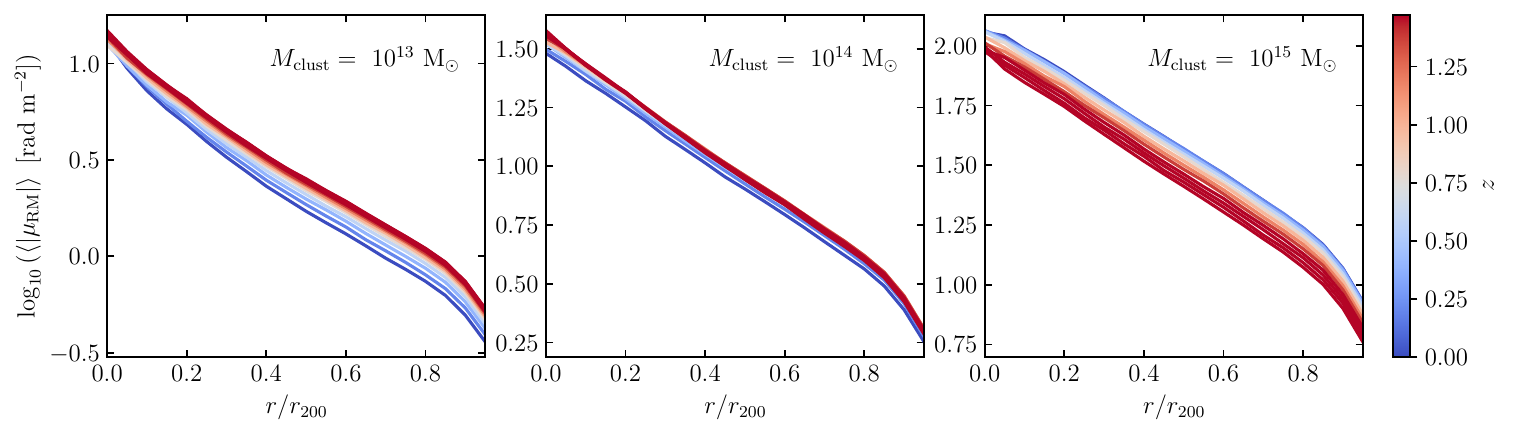}
    \caption{Radial-averaged RM of the equipartition magnetic field $B_{\mathrm{equ}}$ as a function of the radius normalized to $r_{200}$, for the three MT configurations. Each coloured line corresponds to a given redshift value, up to 
    $z = 1.5$.}
    \label{fig:radial_avg_rm_evolution}
\end{figure*}

\begin{figure}
    \centering
    \includegraphics[width = \columnwidth]{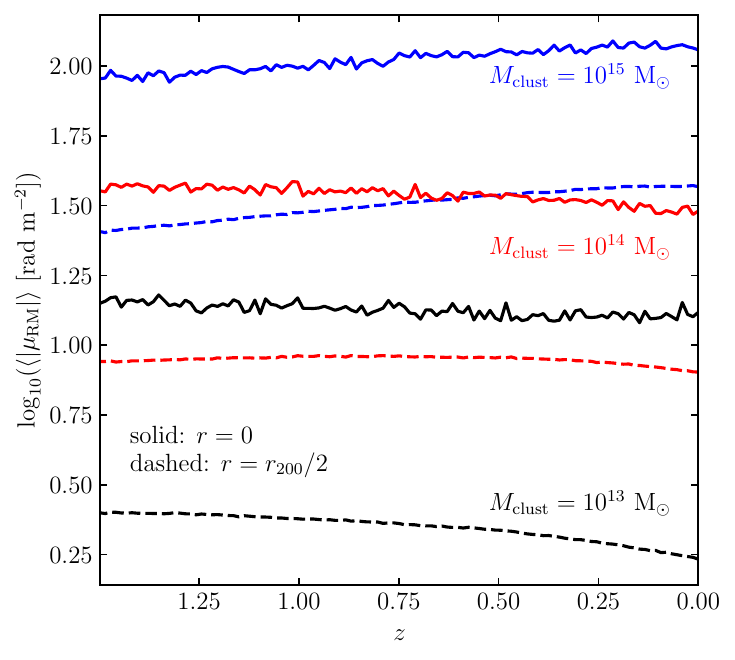}
    \caption{
    Evolution of $\langle |\mu_{\rm RM}|\rangle$
    at $r=0$ (solid lines) and $r= r_{200}/2$ (dashed lines), as a function of the redshift, for the 
    three MT configurations.}
    \label{fig:trend_radial_RM_redshift}
\end{figure}

Figure~\ref{fig:radial_avg_rm_evolution} shows 
RM radial profiles as a function of the redshift of the cluster
for each MT configuration. 
The shape of the profiles does not appear to change over time. However, for $M_{\rm clust}= 10^{13}$ and $10^{14}~\mathrm{M}_{\odot}$, the profiles are slightly shifted upwards as $z$ increases, but the opposite occurs for $M_{\rm clust}= 10^{15}~\mathrm{M}_{\odot}$.
This effect is illustrated in Fig.~\ref{fig:trend_radial_RM_redshift}, which shows the evolution of the value of $\langle|\mu_{\rm RM}|\rangle$ as a function of redshift, for two
values of the radius, 
namely $r=0$ and $r=r_{200}$. There seems to be a mass between $M_{\rm clust}= 10^{14}$ and $10^{15}~\mathrm{M}_{\odot}$ for which the trend reverses.
As shown in Fig.~\ref{fig:plasma_params_evolution}, the evolution of $B_{\rm equ}$ is faster for $M_{\rm clust} = 10^{13}~{\rm M}_{\odot}$ than for the other configurations. 
By associating this with the wavelength dilution given by the factor $(1+z)^{-2}$ in 
Eq.~\eqref{equ:rotation_measure},
as well as the fact that the average RM increases with the mass of the clusters, 
this trend can be explained.
Moreover, the evolution of the equipartition magnetic field is 
intrinsically linked to the evolution of the turbulent velocity field, which in our case is a simplified model. Therefore, there is a possibility that this effect does not originate from 
any underlying physical mechanism, but rather from a 
characteristic inherent to our model.
Testing this hypothesis would require observational data from multiple radio sources in clusters up to $z=1.5$, unavailable today. However, this may prove possible with the next generation of radio telescopes, such as the Square Kilometre Array \citep{heald_2020}.

\begin{figure}
    \centering
    \includegraphics[width = \columnwidth]{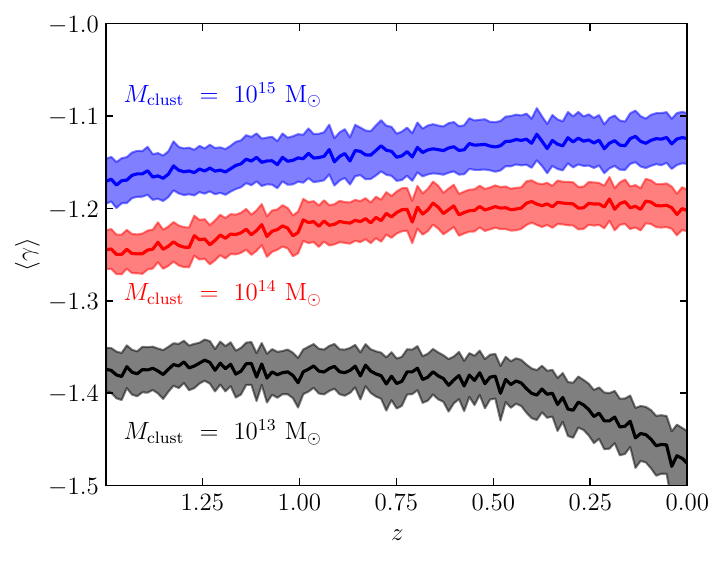}
    \caption{Best-fit values for the fitting parameter $\gamma$ in Eq.~(\ref{eq:equ_radial_rm_lin_fit}). 
    The solid lines represent the skew-normal median of 
    $\gamma$ 
    of all MTs at a given redshift. The shaded areas are the errors given by the skew-normal standard deviation.}
    \label{fig:radial_rm_fitting}
\end{figure}

The profiles in Fig.~\ref{fig:radial_avg_rm_evolution} were fitted with a function of the form
\begin{equation}\label{eq:equ_radial_rm_lin_fit}
\log_{10}\left(\left\langle|\mu_{\mathrm{RM}}|\right\rangle\right) = \gamma \left(\frac{r}{r_{200}}\right)+ \kappa,
\end{equation}
where $\gamma$ and $\kappa$ are fitting parameters. 
The optimal values of $\gamma$ are presented in Fig.~\ref{fig:radial_rm_fitting}. 
Notably, for configurations with $M_{\mathrm{clust}} = 10^{14}$  and $10^{15}~{\rm M}_{\odot}$, 
there is a marginal decrease in the curves as $z \to 0$, but these variations 
are negligible. 
Consequently, it is reasonable to assert that the slope of these profiles remains constant within the examined redshift range. Conversely, for $M_{\mathrm{clust}} = 10^{13}~{\rm M}_{\odot}$, a more substantial temporal variation is observed. Nevertheless, this variation is minor, ranging from approximately 
$\gamma \simeq -1.5$ at $z = 1.5$ to around $\gamma\simeq -1.38$
at $z = 0$. 
However, the discernible difference in trend among these curves cannot definitively be attributed to a specific physical mechanism or potential numerical effects intrinsic to our methods. 

\subsection{Comparison with radio observations}\label{subsec:compare_obs}
In this section we compare the RM radial distribution obtained with our model at $z=0$ 
to various radio observations, to test the robustness of our approach.  
We selected a sample of radio observations of various clusters with masses between $\sim 10^{14}$ and $5\times 10^{15}~{\rm M}_{\odot}$ and redshifts between $z \simeq 0.0140$ and $z \simeq 0.0801$. 
The RM data are retrieved from \citet{bionafede_2010_rot_measure_coma} for A1656, from \citet{govoni_2010_rm_hot_gc} for A119, A514 and A225, and from \citet{clarke_2001_gc_magfields} for A376, A426, A496, A754, A1060, A1314 and A2247.
Relevant data concerning virial mass, virial radius $r_{200}$, and redshift were 
obtained 
from survey catalogues \citep[][and references therein]{girardi_1998_clusters_mass_catalog, 1989_abell_catalog, groener_2016_concentration_mass_cluster}. 
Additionally, we also included radio analyses for A2255 from \citet{govoni_2005_a2255}, and for A400 and A2634 from \citet{eilek_2002_a400_a2634}.
The observed values of the RM correspond to the mean of the normal distribution used to fit the observational data, and the error (when available) corresponds to the standard deviation of the distribution.

\begin{figure*}
    \centering
    \includegraphics[width = \textwidth]{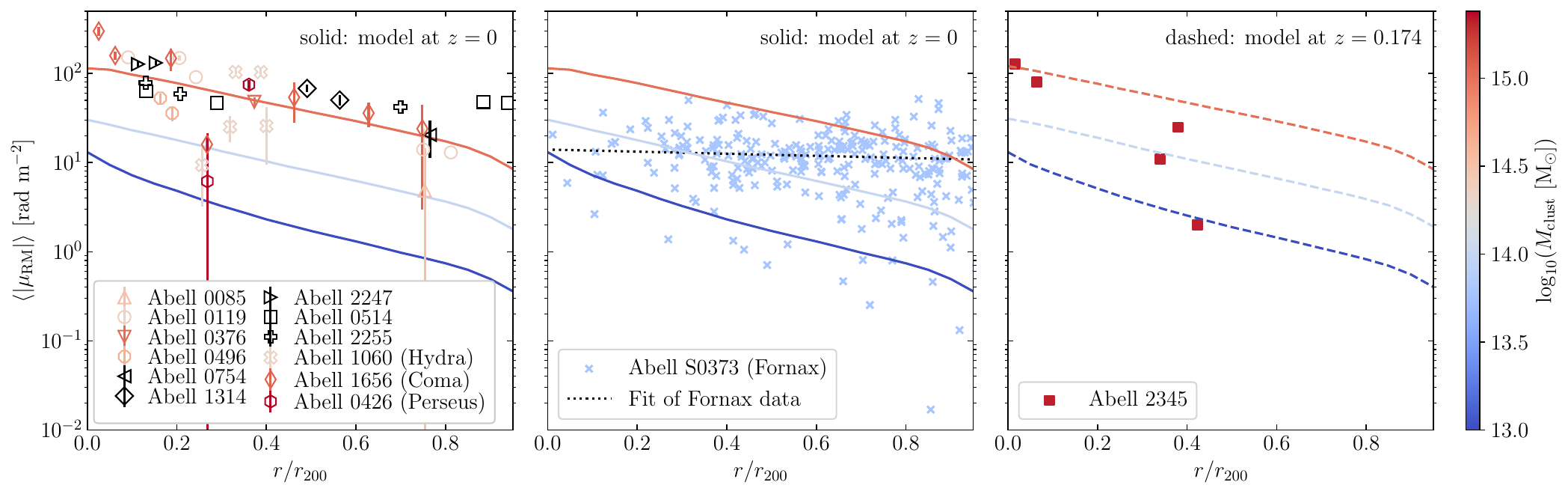}
    \caption{Comparison of the predicted RM radial profiles from our model with radio 
    observations. (\textit{Left}) 
    Comparison of the modelled RM radial profiles with the following observational data from radio sources of a sample of various clusters:
    A1656 (Coma) from \citet{bionafede_2010_rot_measure_coma}; A2255, A119, and A514 from \citep{govoni_2010_rm_hot_gc}; A376, A426 (Perseus), A496, A754, A1060 (Hydra), A1314, and A2247 from \citet{clarke_2001_gc_magfields}; A2255 from \citet{govoni_2005_a2255}; and A400 and A2634 from \citet{eilek_2002_a400_a2634}.
    Our model with $M_{\rm clust} = 10^{15}~\mathrm{M}_{\odot}$ seems to be in good agreement with the observations. In particular, a strong similarity is observed with observations of A1656 (Coma), probably due to its high X-ray emission symmetry. Significant deviations are observed for A1060 (Hydra) and A426 (Perseus), probably due to high AGN activity and strong gas mixing.
    (\textit{Centre}) Comparison of our RM model with the 
    observed data of the AS0373 (Fornax) cluster \citet{anderson_2021_fornax}. 
    The solid lines correspond to our model at $z=0$. The black dotted line corresponds to a fit of the form $10^{\alpha r/r_{200}}$ of the Fornax data, with a best-fit parameter of $\alpha = 0.118\pm0.022$.
    (\textit{Right}) Comparison of our RM model with the observed data of the Abell~2345 cluster from \citet{stuardi_2021_a2345}. 
    The dashed lines correspond to our model at $z = 0.174$, which is the closest redshift of our model to that of Abell~2345, which is estimated roughly at $z\simeq 0.0179$. Both observations are in relatively good agreement with our model up to $r/r_{200} \simeq 0.4$. Above this limit, the discrepancies are most likely due to the dynamic state of both clusters, which are supposed to be in an ongoing-merger state.}
    \label{fig:radial_avg_rm_observations}
\end{figure*}

The comparison of our model with such observations is shown in the left-hand panel of Fig.~\ref{fig:radial_avg_rm_observations}.
The solid lines correspond to the radial profiles of our model at $z=0$ for the three MT configurations.
The colours indicate the observed 
cluster mass or the value $M_{\rm clust}$ in our model, respectively. 
The black markers indicate radio observations for clusters whose mass has not been estimated. 
Despite obvious variations between our model and the reported observational 
values due to many factors discussed in the next section, the overall trend 
of the RM as a function of radius appears to be in good agreement with 
observations for $M_{\rm clust} = 10^{15}~\mathrm{M}_{\odot}$, which is an encouraging 
indication of the robustness of our approach. 
It should be noted, however, that our model and the observational data do not 
represent the same entities. 
The curves in our model represent the average of all RM values at a given radius. 
The observational data points reported represent the average RM value observed for a radio source with a certain spatial extent, at a given distance from the cluster's centre. 

We also compared our model with the detailed observational data of a low-mass and low-redshift cluster, namely Abell~S0373 (Fornax). 
The polarimetric data of Fornax sources are retrieved from \citet{anderson_2021_fornax}. We compute the projected angular distance from NGC~1399 (considered in the original paper as the centre of the cluster) assuming that the galaxy is located at a distance of 20.24~Mpc \citep[e.g.][]{lavaux_2011_redshift_catalog}. Fornax's virial radius is assumed to be $r_{200} = 0.7$~Mpc, its redshift 
$z = 0.005$ and its mass $M = 6\times 10^{13}~\mathrm{M}_{\odot}$ \citep[e.g.][and references therein.]{maddox_2019_fornax, anderson_2021_fornax}.
The results are shown in the middle panel of Fig.~\ref{fig:radial_avg_rm_observations} respectively. We have plotted the curves corresponding to our model at $z=0$ (solid lines).
The Fornax data were fitted with a function proportional to $10^{\alpha r/r_{200}}$ where $\alpha$ is a real number, which corresponds to the black dotted line in Fig.~\ref{fig:radial_avg_rm_observations}. It appears that our model with 
$M_{\rm clust} = 10^{14}~{\rm M}_{\odot}$ and the observations from \citet{anderson_2021_fornax} are in relatively good agreement up to about $r/r_{200} \simeq 0.4$, and for $\alpha \simeq -0.118\pm0.022$. However, the RM profile of our model decreases, while the Fornax profile seems to be almost constant with $r$. However, Fornax is supposed to be in an ongoing merger with the Fornax~A substructure, which tends to make the assumption of spherical symmetry of the baryonic gas inapplicable for this cluster. Nevertheless, the good match between the central region of Fornax and our model is an encouraging sign of the robustness of our approach and seems to confirm that lower-mass clusters produce weaker RM signals than their more massive counterparts.

Finally, we also compared our model with the observational data of a massive, high-redshift cluster, namely Abell~2345. Its various data are retrieved from \citet{stuardi_2021_a2345}. 
Its redshift is reported as $z=0.1789$, its mass as $M =2.85 \times 10^{15}~{\rm M}_{\odot}$, and its virial radius as $r_{200}=3.4~\mathrm{Mpc}$ \citep[all comoving, see e.g.][]{boschin_2010_a2345_structure}. The results are shown on the right panel of Fig.~\ref{fig:radial_avg_rm_observations}. We have plotted the curves corresponding to our model at $z=0.174$ (dashed lines), which is the closest redshift of our model to that of Abell~2345. Its central part also seems to be in good agreement with our results. Similarly, this cluster presents two non-symmetric radio relics, and \citet{boschin_2010_a2345_structure} have suggested that it is undergoing a merger, making its matter distribution highly asymmetric as well, which tends to explain the discrepancies observed at $r/r_{200} \simeq 0.4$. It should also be noted that the higher redshift value of Abell~2345 does not alter our predictions regarding its RM values, as shown by the various curves of our model in the right-hand panel of Fig.~\ref{fig:radial_avg_rm_observations}.
In addition, several obvious aspects that could change our results are discussed in 
Sect.~\ref{sec:discussions}.

\section{Discussion}\label{sec:discussions}
\subsection{Missing physics}
The advantage of our model is its simplicity which allows us to scan a large range of characteristic parameters of galaxy clusters. 
However, our model is relatively basic, and several factors might influence our results. 
Firstly, our assumptions of spherical symmetry and the nature of plasma are 
rough. 
Although we aim to study global RM trends, it is well known that the ICM is multiphase. 
For instance, observations have shown the presence of regions of cold neutral hydrogen gas, calling into question the dominant character of hot gas in the ICM \citep{bonamente_2001_cold_hot_gas}.
The effect of the multiphase nature of the ICM on X-ray observables was investigated using data from hydrodynamic simulations, including components such as AGN feedback \citep{zuhone_2023_multiphase_xray}. Furthermore, from recent turbulent magnetohydrodynamic simulations, it is known that the multiphase nature of the medium affects magnetic field properties 
\citep{seta_federrath_2022, mohapatra_2022, das_gronke_2024}. We also adopted a very simple turbulent velocity model, assuming that turbulence was only generated by shocks and shear from mergers of dark matter halos. However, recent 3D MHD simulations have shown, for example, that turbulence in the ICM hot gas could also have contributions from AGN jets or the stream of precipitating cold filaments \citep{wang_2021_turb_ICM}.

Finally, we assume that the SSD amplifies the magnetic field 
up to equipartition with the turbulent velocity field and that 
the magnetic energy peak shifts towards
larger spatial scales in the nonlinear dynamo phase. At saturation, we assume that the peak of the magnetic energy spectrum has reached the turbulent forcing scale. 
While such a shift of magnetic energy towards larger scales is observed in numerical simulations \citep[e.g.][]{Federrath_2014_turbdyn, seta_2021_dyn}, it remains unclear whether this shift continues up to the forcing scale.
Moreover, this is even less known and understood for weakly collisional plasmas.

\subsection{Possible observational effects}
We have seen that our predictions for the RM are in good agreement with observations of multiple radio sources in a set of about ten clusters. However, several factors could bias our results. 
Our RM maps are calculated on the assumption that each numerical cell is a source of linearly polarized wave emission. 
In this case, the redshift of the source corresponds 
to the redshift of the cluster studied. 
Depending on the redshift of the emitting source, this may, for example, cause a vertical shift in the observations. 
The effect of intracluster sources is also not analysed, nor is the emission of cosmic ray electrons from the ICM. 
However, in observations presented in \citet{bionafede_2010_rot_measure_coma} some radio sources are located within the virial radius of the cluster. Also, it is common for several distinctive areas of RM emission to originate from the same radio source. Depending on the reverse scale of the ICM's magnetic field, some of these zones may thus present different values, sometimes of opposite sign, producing a discontinuous RM distribution \citep[e.g.][]{govoni_2010_rm_hot_gc, bionafede_2010_rot_measure_coma}. Averaging such data using a Gaussian fit could produce a different standard deviation value from that observed if more data were available. More generally, the cosmic ray and thermal electrons in the ICM are mixed and thus the emission and Faraday rotation of the emission happens within the same region. This might affect some of our results, especially the estimates of $\sigma_{\rm RM}$. 
We also assumed that the resolution of our RM maps was set by the turbulent driving scale, as the SSD predicts that the power spectrum of magnetic energy shifts towards such length scales from the moment the magnetic energy density is more or less in equipartition with the kinetic energy density of the turbulence. However, the non-linear effects of dynamo in WCPs have yet to be fully elucidated. Thus, if the typical length scale of the magnetic field were to vary spatially, the results could differ from our predictions.

\section{Conclusions}\label{sec:conclusions}

Magnetic fields in the microgauss range are systematically observed in the ICM of galaxy clusters. 
The previous work by \citet{rappaz_schober_2024}, 
which was motivated by the pioneering work of \citet{Schekochihin_2006_magfield_gclust}, \citet{st-onge_2018_hybrid}, and \citet{st-onge_kunz_squire_scheko_20202},
suggests that if such fields were amplified from primordial fields, the growth of the magnetized field would be much faster than suggested by classical MHD theory. 
In this paper, we study the evolution of the RM implemented in MTs generated by the modified GALFORM algorithm \citep{parkinson_2008_mod_galform}, assuming that the magnetic field had already been amplified to equipartition with the turbulent velocity field.
The main outcomes of this work are the following.

The average RM is the highest for our MT configuration with $M_{\rm clust} = 10^{15}~\mathrm{M}_{\odot}$, and the lowest with $M_{\rm clust} = 10^{13}~\mathrm{M}_{\odot}$. 
In all MT configurations, the mean RM amplitude does not change significantly up to redshift $z \simeq 1.5$. This effect is due to two contributions. Between $z_{\rm max} \simeq 1.5$ and $z=0$, the mean thermal electron density $n_e$ decreases, which causes the RM to be higher at $z_{\rm max}$. At the same time, the RM depends on the redshift as $(1+z)^{-2}$.
The two effects 
compensate
for each other, resulting in no 
major evolution
in the RM.                                     
For $M_{\rm clust} = 10^{14}$ and $10^{15}~\mathrm{M}_{\odot}$, 
the RM radial profiles $\mu_{\rm r, RM}$ follow approximately the same power law, 
that is,~$\log_{10}(\mu_{\rm r, RM}) \propto -1.2(r/r_{200})$. 
For $M_{\rm clust} = 10^{13}~\mathrm{M}_{\odot}$, this relation is $\log_{10}(\mu_{\rm r, RM}) \propto -1.4(r/r_{200})$.
Overall, the radial profile associated with our model with $M_{\rm clust} = 10^{15}~\mathrm{M}_{\odot}$ 
to $z=0$ is in good agreement with radio observations from various clusters. 
The observations of Abell1656 (Coma) are the closest to our model. 
We assume that the main reason for such a good agreement is 
that
Coma has recently (approximately 1 Gyr ago) undergone a major merger event with the 
NGC4839 group,
and is now settling in a configuration close to hydrostatic equilibrium, as suggested by \citet{neumann_2003_coma_state} and \citet{churazov_2021_erosita}.
This gives it a symmetry in X-ray emission that is very close to our assumption of 
spherical symmetry for the plasma quantities implemented in our subhalos. 
On the other hand, our model does not agree well with certain values of the Abell426 
(Perseus) observations. 
We assume that this is due to strong AGN activity in Perseus' core and in particular 
an X-ray emission structure suggestive of strong gas 
mixing \citep[see e.g.][]{zhuravleva_2014_turb_heating}.

Our model does not include certain factors likely to modify the observed RM 
value, such as AGN jets, the multiphase nature of the ICM, 
and sources of turbulence besides merging halos.
Despite this, we suggest that our model can be used to efficiently create radial 
RM distributions for redshifts up to $z\simeq 1.5$, for clusters with masses of 
the order of $10^{15}~\mathrm{M}_{\odot}$.
The complete analysis, from the raw data of the modified GALFORM to the 
results presented in this work, does indeed take just one or two days, on a set of around ten conventional CPUs. 
This is very time-saving compared with full-scale MHD cosmological simulations, 
which can take several months or even years to complete and process. 
Finally, our model could prove useful in the creation of mock radio observations 
for next-generation radio telescopes, such as the Square Kilometre Array \citep{heald_2020}.

\begin{acknowledgements}
We are grateful to Jean-Paul Kneib, Antoine Rocher, and Aurélien Verdier
for very useful discussions on the cosmology part of this work. We thank Radhika Achikanath Chirakkara for useful discussions related to dynamos in galaxy clusters.
YR and JS acknowledge the support from the Swiss 
National Science Foundation under Grant No.\ 185863.
CF~acknowledges funding provided by the Australian 
Research Council (Discovery Project DP230102280), and the 
Australia-Germany Joint Research Cooperation Scheme (UA-DAAD).
\end{acknowledgements}


\begin{appendix}
\section{RM maps resolution}\label{app:resolution_rm_maps}
\begin{figure}
    \centering
    \includegraphics[width = \columnwidth]{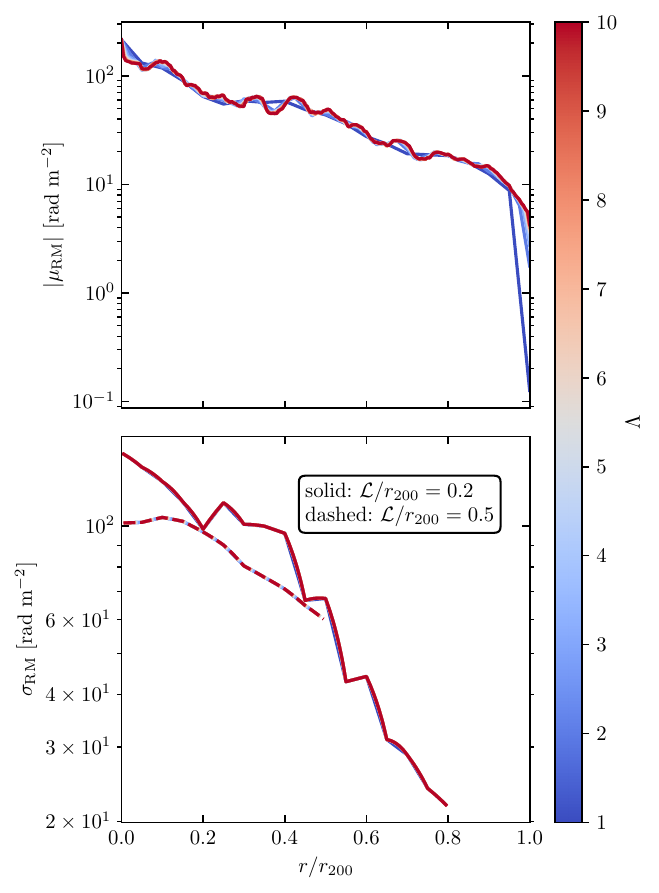}
    \caption{Effect of increasing the resolution in each grid cell of an RM map by a factor of $\Lambda$, on the radial profile of the RM (top) and its standard deviation (bottom). The different curves are calculated for a single tree with $M_{\rm clust} = 10^{15}~\mathrm{M}_{\odot}$.}
    \label{fig:effect_incr_factors}
\end{figure}

In this section we study the effect of both numerical resolution and forcing scale values on the computed signals of the RM. Figure \ref{fig:effect_incr_factors} shows the effect of increasing the resolution in each grid cell of an RM map by a factor $\Lambda$ on the radial profile of the RM and its standard deviation. In particular, the profiles are computed for a single realization of a merger tree with $M_{\mathrm{clust}} = 10^{15}~\mathrm{M}_{\odot}$, at $z=0$. For $\sigma_{\rm RM}$, we compute the profiles for $\mathcal{L}/r_{200}=0.2$ and $0.5$. Overall, there is no difference between the curves, and we directly conclude that $\Lambda$ does not affect the final result.

\section{Effect of different averages}\label{appendix:different_avgs}
\begin{figure}
    \centering
    \includegraphics[width = \columnwidth]{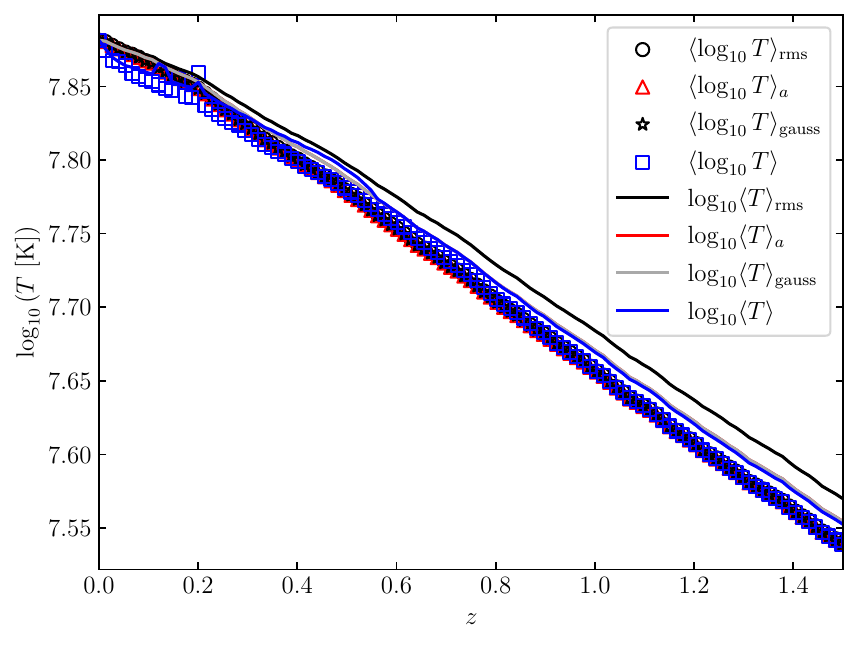}
    \caption{Evolution of the temperature $T$ for $M_{\mathrm{clust}} = 10^{15}~\mathrm{M}_{\odot}$ and $z_{\mathrm{max}}$ calculated in different ways. The lines correspond to the arithmetic mean $\langle\rangle_a$, the root-mean-square $\langle\rangle_{\rm rms}$, the median of a skew-normal fitting $\langle\rangle$, and the average of a Gaussian fitting $\langle\rangle_{\rm gauss}$, for which the logarithm was taken. The markers correspond to the same average processes but are performed on the logarithmic values of $T$.}
    \label{fig:effect_different_avgs}
\end{figure}

Here, we present an analysis of the effect of different types of averages on the evolution of the temperature, for our model with $M_{\mathrm{clust}} = 10^{15}~\mathrm{M}_{\odot}$. To do this, we calculated the temperature profile for each $N_{\mathrm{tree}}=10^3$ MTs. At each fixed redshift value, we estimated the mean of the temperature values of all MTs in different ways. Specifically, we calculated the arithmetic mean $\langle T\rangle$, the root mean square $\langle T\rangle_{\mathrm{rms}}$, the median value $\langle T\rangle_{\mathrm{SN}}$ of the fit of the data with a skew-normal distribution, as well as the mean $\langle T\rangle_{\mathrm{gauss}}$ of the fit of the data with a normal distribution. We then repeated the same operation but on logarithmic temperature values. 
Figure~\ref{fig:effect_different_avgs} shows the different results of such averaging processes on temperature evolution. It is clear that the variations between averages are minor, and that the arbitrary choice of one of these processes does not change the results and conclusions of our study.
\end{appendix}

\end{document}